\documentclass[utf8]{frontiersFPHY} 

\usepackage{url,hyperref,lineno,microtype,subcaption}
\usepackage[onehalfspacing]{setspace}

\newcommand{\gsim}{\lower.7ex\hbox{$\;\stackrel{\textstyle>}{\sim}\;$}}
\newcommand{\lsim}{\lower.7ex\hbox{$\;\stackrel{\textstyle<}{\sim}\;$}}
\newcommand{\TeV}{\rm{TeV}}

\def\firstAuthorLast{Abdelhak Djouadi {et~al.}}
\def\Authors{\mbox{A. Djouadi$^{1,2}$, J.C. Criado$^3$, N. Koivunen$^2$, K. M\"{u}\"{u}rsepp$^2$, M. Raidal$^2$ and H. Veerm\"{a}e$^2$}
}
\def\Address{
$^1$ CAFPE and Departamento de F\'isica Te\'orica y del Cosmos, Universidad de Granada, E-18071 Granada, Spain.\\
$^2$ Laboratory of High Energy and Computational Physics, NICPB, R\"avala pst. 10, 10143 Tallinn, Estonia. \\
$^3$ Institute for Particle Physics Phenomenology, Department of Physics, Durham University, Durham DH1 3LE, United Kingdom. 
}
\def\corrAuthor{Abdelhak Djouadi} 
\def\corrEmail{adjouadi@ugr.es}

\newcommand{\be}{\begin{equation}}
\newcommand{\ee}{\end{equation}}
\newcommand{\bea}{\begin{equation}\begin{aligned}}
\newcommand{\eea}{\end{aligned}\end{equation}}

\begin{document}
\onecolumn
\firstpage{1}

\title[New fermions and the g--2]{\mbox{New fermions in the light of the $\mathbf{(g-2)_\mu}$}} 

\author[\firstAuthorLast ]{\Authors} %This field will be automatically populated
\address{} %This field will be automatically populated
\correspondance{} %This field will be automatically populated

\extraAuth{}

\maketitle

\begin{abstract}

\section{}
\vspace*{-2mm}

The very precise measurement of the anomalous magnetic moment of the muon, recently released by the Muon g-2 experiment at Fermilab, can serve to set stringent constraints on new particles. If the observed 4$\sigma$ discrepancy from the Standard Model value is indeed real, it will set a tight margin on the scale of the masses and couplings of these particles. Instead, if the discrepancy is simply a result of additional theoretical and experimental uncertainties to be included, strong constraints can be put on their parameters. In this mini-review, we summarize the impact of the latest muon g-2 measurement on new fermions that are predicted by a wide range of new physics models and with exotic quantum numbers and interactions. We will particularly discuss the case of vector-like leptons, excited leptons, and supersymmetric fermions, as well as spin-3/2 isosinglet fermions, which have been advocated recently. 
\vspace*{-2mm}
\end{abstract}

\section{Introduction}
\vspace*{-1mm}

The Fermilab Muon $g-2$ collaboration has recently released~\cite{Abi:2021gix} a new measurement of the anomalous magnetic moment, $a_\mu= \frac12 (g-2)_\mu$, of the muon, 
$a_\mu^{\rm Fermilab}= (116 592 040 \pm 54)\times10^{-11}$,
representing a $3.3 \sigma$ deviation from the Standard Model (SM) value, for which a wide consensus among theorists gave the prediction
\cite{Aoyama:2020ynm}
\be
 a_\mu^{\rm SM}= (116 591 810 \pm 43)\times10^{-11}\, ,
\ee
before a new lattice QCD analysis \cite{Borsanyi:2021} predicted a value that is more agreeing with the SM expectation. When combined with the result of the previous Brookhaven muon experiment \cite{Bennett:2006fi} which had a deviation of about $3.7\sigma$ from the SM expectation, one obtains a final result
\be 
 a_\mu^{\rm EXP}= (116 592 061 \pm 41)\times10^{-11} \, ,
\ee
which implies a $4.2 \sigma$ deviation from the SM prediction (if the new lattice result is ignored) \cite{Abi:2021gix} 
\be
 \Delta a_\mu= a_\mu^{\rm EXP} - a_\mu^{\rm SM} = (251 \pm 59)\times10^{-11} \, .
\ee
It is extremely tempting to attribute the discrepancy $ \Delta a_\mu$ to additional contributions from models of new physics beyond the SM and, before the issue of the theoretical uncertainties is resolved, this is the attitude that we choose to take. In any case, if the discrepancy is alleviated or eliminated by a more refined theoretical description, the new measurement would allow to strongly constrain the basic parameters of this new physics and in a way that should be complementary to the direct searches that are performed in the high-energy frontier experiments at the Large Hadron Collider (LHC). 

In this review note, we will confront this new and precise $(g-2)_\mu$ result with the predictions coming from a variety of models beyond the SM, which contain additional heavy fermions. These particles can have the usual lepton and baryon quantum numbers but come with exotic ${\rm SU(2)_L \times U(1)_Y}$ assignments.

A well-known example of such a possibility is given by vector-like fermions, when both the left- and right-handed components appear in the same electroweak doublet, allowing for a consistent generation of their masses without the need of the Higgs mechanism. These fermions often occur in grand unified theories \cite{Djouadi:1995es} and have been advocated e.g. to explain the hierarchies in the SM flavour sector \cite{Giudice:2012ms,Kannike:2011ng,Dermisek:2013gta}. One can also have sequential fermions, such as a fourth generation, or mirror fermions which have chiral properties that are opposite to those of the SM fermions. However, it is necessary to modify the SM Higgs sector in order to evade the strong constraints from the precise determination of the Higgs boson properties at the LHC~\cite{Djouadi:1997rj,Denner:2011vt,Djouadi:2012ae,Kuflik:2012ai}. The mixing of the heavy and light fermions that have the same U(1)$_{\rm Q}$ and SU(3)$_{\rm C}$ quantum numbers gives rise to new interactions~\cite{Djouadi:1995es,Djouadi:1993pe} which allow for the decays of the heavy states into the lighter ones and to generate contributions which could be observed in highly precise experiments. 

Another type of new fermions which have been discussed in the past are excited fermions. They are a characteristic signature of compositeness in the matter sector which was and is still advocated to explain some pattern in the mass spectrum. The SM fermions would then correspond to the ground states of the spectrum and the excited states would decay to the former ones through a magnetic type de-excitation. In the simplest case, the excited fermions have spin and isospin $\frac12$, and the transition between excited and fundamental fermions is described by an ${\rm SU(3)_C\times SU(2)_L\times U(1)_Y}$ invariant effective interaction of the magnetic type~\cite{Djouadi:1995es,Boudjema:1992em}. Hence, besides the full-strength couplings to gauge bosons, excited states have couplings to SM fermions and gauge bosons that are inversely proportional to the compositeness scale $\Lambda$. These couplings determine the decay and production properties of the excited states and, e.g. induce anomalous contributions to the dipole moments. 

We will also discuss the case of supersymmetric theories in their minimal version, the so-called minimal supersymmetric extension of the SM or MSSM. In this scenario, the Higgs sector is enlarged to contain two doublet fields and each SM particle or additional Higgs boson has a supersymmetric partner with a spin that differs by $\frac12$. The superpartners of the gauge and Higgs bosons will mix to give the physical states, the spin-$\frac12$ charginos and neutralinos, with the lightest neutralino being the lightest SUSY particle which is stable and forms the dark matter. The charginos and neutralinos would contribute to the $(g-2)_\mu$ along with the scalar partners of the muon, the smuons and their associated sneutrinos. These particles have been, for a long time, considered as the best candidates to explain the previous discrepancy in the measurement.

Finally, we will also discuss new particles with a spin higher than unity and, in particular, we will consider the case of a massive electrically neutral and colourless spin-$\frac32$ fermion, which was recently discussed in dark matter \cite{Criado:2020jkp}, collider \cite{Criado:2021itq} and nuclear physics \cite{Criado:2021gcb} phenomenology but also in the context of the new $g-2$ value \cite{Criado:2021qpd}. Massive spin-$\frac32$ particles are present in supersymmetric extensions of gravity and string-theoretical models. The phenomenological studies of generic higher-spin particles had severe problems in the past, related to the non-physical degrees of freedom in their representations that need to be eliminated as they lead to pathologies like the violation of causality and perturbative unitarity. These problems are avoided in a recently proposed Effective Field Theory (EFT) approach to generic higher-spin particles~\cite{Criado:2020jkp,Criado:2021itq,Criado:2021gcb} which considers only the physical degrees of freedom. Spin-$\frac32$ fermions have interactions with leptons and hence generate a contribution to the muon $(g-2)$. We will summarize it here and compare it with the corresponding results for the spin-$\frac12$ fermions.

%§§§§§§§§§§§§§§§§§§§§§§§§§§§§§§§§§§§§§§§§§§§§§§§§§§§§§§§§§§§§§§§§§§§§§§§§§§§§§§§§§§§

\section{New fermion contributions to the g--2}

\subsection{Vector-like leptons}

For charged heavy leptons with exotic ${\rm SU(2)_L \times U(1)_Y}$ quantum numbers, except for singlet heavy neutrinos without electromagnetic or weak charges, the couplings to the photon, the $W$ and the $Z$ bosons are unsuppressed. The heavy states mix with the SM leptons in a model-dependent and a possibly rather complicated manner, especially if different fermion generations can mix. 

In the following, we will consider as an example the case of vector-like leptons that have been introduced in order to explain flavour hierarchies in the SM; see Refs.~\cite{Giudice:2012ms,Kannike:2011ng,Dermisek:2013gta} for detailed studies. Two doublets $L_L$ and $L_R$ and two singlets $E_L$ and $E_R$ are introduced with a Lagrangian given by \cite{Giudice:2012ms}
\be \label{eq:lagrangian}
 { L} \propto M_E \bar{E}_{L} E_{R} + M_L \bar{L}_{R} L_{L} + m_E \bar{E}_{L} e_{R} +
m_L \bar{L}_{R} \ell_{L} + \lambda_{LE} \bar{L}_{L} E_{R} \Phi + {\bar\lambda}_{LE} \bar{L}_{R} E_{L} \Phi^\dagger
\!+\! {\rm h.c.},
\ee
with the $L_L, E_R$ and $L_R,E_L$ fields having, respectively, the same and opposite quantum numbers as the SM leptons $\ell_L, e_R$; $\Phi$ is the SM Higgs doublet. The mass eigenstates are obtained by diagonalizing the mass mixing in ${L}$ through $2\! \times\! 2$ unitary matrices, where the mixing angles read $\tan\theta_L = m_L/M_L$ and $\tan\theta_R = m_E/M_E$. After rotating the fields, the previous Lagrangian becomes 
\bea
    {L} & \propto \sqrt{M^2_E + m^2_E}\, \bar{E}_L E_R + \sqrt{M^2_L+m^2_L}\, \bar{L}_R L_L + {\bar\lambda}_{LE}\, \bar{L}_R E_L \Phi^\dagger \\
    &+ \lambda_{LE}\, \left( \sin {\theta_L} \sin {\theta_R} \bar{\ell}_L e_R  + 
    \cos {\theta_L} \sin{\theta_R} \bar{L}_L e_R \!+\! \sin {\theta_L} \cos{\theta_R} \bar{\ell}_LE_R \! +\! \cos {\theta_L} \cos{\theta_R}~\bar{L}_L E_R \right) \Phi + {\rm h.c.}\,,
\eea
After symmetry breaking, the spectrum will consist of two heavy leptons with masses $\sqrt{M^2_L + m^2_L}$ and $\sqrt{M^2_E + m^2_E}$ and the light leptons with masses given approximately by $m_{\ell_i} \simeq \lambda_{LE}\, \sin_{\theta_L}^i \sin_{\theta_R}^i v+ \mathcal{O}(v^2/ M^2_{L,E})$, where $i$ is the generation index and $v\simeq 246$ GeV the Higgs vev. Notice that the Yukawa couplings have been assumed to be zero and are generated after electroweak symmetry breaking through the mixing between heavy and light fermions, once the former have been integrated out. 

The heavy charged and neutral leptons contribute to the anomalous magnetic moment through Feynman diagrams that involve the exchange of two $W$ bosons with the neutral lepton and the exchange of two charged states with a $Z$ or Higgs boson. Heavy exotic fermion contributions to leptonic $(g-2)$ have been also discussed and evaluated in Refs.~\cite{Kannike:2011ng,Dermisek:2013gta,Giudice:2012ms,Rizzo:1985db,Nardi:1991rg,Vendramin:1988zc,Kephart:2001iu,Csikor:1991np,Chavez:2006he,Aboubrahim:2016xuz,Crivellin:2018qmi}. Here,we simply display the contributions to $a_\mu$ in the limit of small mixing angles, retaining only terms of order $v^2 / M^2_{L,E}$
\be
    \Delta a_\mu \simeq \frac{1}{16\pi^2}\frac{m^2_\mu}{M_L M_E} {\rm Re}(\lambda_{LE}{\bar\lambda_{LE}}) \approx  10^{-9}~{\rm Re}(\lambda_{LE}{\bar\lambda_{LE}})\,
\left(\frac{300~{\rm GeV}}{\sqrt{M_L M_E}}\right)^2.
\ee
Thus, for $M_L,M_E$ values of the order of the electroweak symmetry breaking scale $v$ and for large Yukawa couplings to the muon $\lambda_{LE}, \bar\lambda_{LE}$, the contributions to $a_\mu$ can be significant.

\subsection{Excited leptons}

In the case of the charged excited leptons that we will denote by $\ell^\star$, we assume for simplicity that they have spin and isospin $\frac12$. Besides the $\ell^\star \ell^\star V$ interaction with the $V=\gamma, W, Z$, gauge bosons, there is a magnetic-type coupling between the excited leptons, the ordinary ones and the gauge bosons $\ell^\star \ell V$ which allows for the decays of the heavy states, $\ell^\star \to V\ell$ \cite{Boudjema:1992em}. This coupling induces a contribution to the anomalous magnetic moment of the lepton via diagrams in which there is a transition of the magnetic type with a $\mu^\star \mu^\star$ loop along with the $Z$ boson and the photon, as well as diagrams in which the magnetic transition occurs at the $\gamma \mu^\star \mu$ vertex. The Lagrangian describing this transition should respect a chiral symmetry in order to induce not excessively large contributions to the anomalous moment. As a consequence, only the left- or the right-handed component of the excited lepton takes part in the generalized magnetic d -excitation. The corresponding Lagrangian then reads 
\be
    {L}_{ \ell \ell^{\star} \gamma} 
    = \frac{e \kappa_{L/R} }{\sqrt 2 \Lambda} \bar{\ell^\star} \sigma^{\mu \nu} \ell_{L/R} F_{\mu \nu}+ {\rm h.c.}\ .
\ee
This interaction could be generalized to the ${\rm SU(2)_L \times U_Y(1)}$ case where the photon field
strength is extended to the $W_{\mu\nu}$ and $B_{\mu \nu}$ ones. In the equation given above, 
$\Lambda$ is the compositeness scale that we set to 1 TeV. We will set all the weight factors for the field strengths to $\kappa_{L/R}$ to simplify the analysis and ensure that the excited neutrino has no tree-level electromagnetic couplings~\cite{Boudjema:1992em}. Thus, apart from the masses of the excited leptons that we will also equate, $m_{\ell^\star}= m_{\nu^\star_\ell}$, the only free parameter will be the strength $\kappa_{L,R}/\Lambda$ of the de-excitation which involves either a left-handed or a right-handed fermion.

The contribution $\Delta a_\mu$ of the $\mu^\star$ and its partner $\nu^\star_\mu$ to the muon magnetic moment has been calculated long ago \cite{Renard:1982ij,delAguila:1984sw,Choudhury:1984bu,Mery:1989dx,Rakshit:2001xs} and the result in the case where the simplifications above are performed, assuming $m_{\ell^\star}= m_{\nu^\star_\ell}= \Lambda \gg M_W$, which we believe to be a good approximation, is simply given by 
\cite{Mery:1989dx}
\be
    \Delta a_\mu = \frac{\alpha}{\pi} \frac{\kappa_{L,R}^2}{\Lambda^2} m_\mu^2 c_{L/R},
\ee
where the numerical values of the $c_L,c_R$ coefficients in these limits are $c_L \simeq 10$ and $c_R \simeq 5.3$, respectively for left-handed $V \mu^\star \mu_L$ and right-handed $V \mu^\star \mu_R$ transitions. 

\subsection{Supersymmetric particles}

In this subsection, we will briefly discuss the contributions to $a_\mu$ of the superparticles in the minimal supersymmetric extension of the SM (MSSM) \cite{Drees:2004jm}, namely the one with the chargino-sneutrino and neutralino-smuon loops. These have also been calculated long ago  \cite{Ellis:1982by,Grifols:1982vx,Barbieri:1982aj,Kosower:1983yw, Chattopadhyay:2001vx,Carena:1996qa,Martin:2001st,Chakraborti:2021kkr,Moroi:1995yh} and the approximate result, taking into account only the chargino-sneutrino contribution which is an order of magnitude larger than the one of the neutralino-smuon loop, is rather simple and accurate \cite{Moroi:1995yh} 
\be
    \Delta a_\mu 
    \simeq \frac{\alpha}{8 \pi s_W^2} \tan\beta \times \frac{m_\mu^2} {\tilde m^2} 
    \approx 1.5 \times 10^{-11} \tan\beta \left[ \frac{\tilde m}{\TeV} \right]^{-2},
\ee
where $\tan\beta$ is the ratio of vacuum expectation values of the two doublet Higgs fields that break the electroweak symmetry, $1 \lsim \tan\beta \lsim  60$ and $\tilde m$ is a SUSY scale given by the largest mass between the chargino and the sneutrino states, $\tilde m = {\rm max} (m_{\tilde \nu}, m_{\chi_1^+})$. Hence, a large SUSY contribution to $a_\mu$ can be generated for large enough $\tan\beta$ values and superparticle masses of order a few hundred GeV. 

We note that the sign of the SUSY contribution is equal to the sign of the higgsino mass parameter $\mu$, $\Delta a_\mu \propto (\alpha/\pi) \times \tan\beta (\mu M_2)/ \tilde m^4$ with $M_2$ the gaugino (wino) mass parameter.
On should note too that the extended two-Higgs doublet Higgs sector of the MSSM will in principle also contribute to the $g-2$, but as the Higgs particles are heavy or do not have strong couplings to muons, the impact is expected to be very small. This might not be the case in extensions of the MSSM, such as the next-to-minimal supersymmetric SM or NMSSM, in which one could have a light pseudoscalar Higgs boson with enhanced couplings to muons \cite{Djouadi:2008uw} that could significantly contribute to the $g-2$ \cite{Arcadi:2021zdk}. 

\subsection{Spin-3/2 fermions}

Among the dimension-7 operators which describe the interactions with the SM fields of a charge and colour neutral SM isosinglet spin-$\frac32$ field denoted by $\psi_{3/2}$~\cite{Criado:2021itq}, the following ones will contribute to the $g-2$, keeping only couplings to (second generation) leptons 
\be 
-\mathcal{H} =	\frac{1}{\Lambda^3} \psi_{3/2}^{abc} \Big[ 
	c_B \tilde{\phi}^\dagger \sigma^{\mu\nu}_{ab} B_{\mu\nu} L_{Lc}^2
+	c_W \tilde{\phi}^\dagger \sigma^{\mu\nu}_{ab} \sigma_n W^n_{\mu\nu} L_{Lc}^2 
\ \Big] 
   + \text{h.c.},
\ee
where $a,b,c$ are two-spinor indices; $L^i_a$ are the left-handed lepton doublets $L^i_{La} = (\nu_{La}^i, e_{La}^i)$; $B_{\mu \nu}$ and $W_{\mu \nu}$ denote the ${\rm U(1)_Y}$ and ${\rm SU(2)_L}$ field strengths; and $\phi$ is the SM Higgs doublet given in the unitary gauge by $\phi = (0,H+v)/\sqrt{2}$ (with $v=246$ GeV and $H$ the SM Higgs boson). The constant tensors $\sigma^\mu_{a\dot{a}}$ are given in terms of the identity matrix $\sigma^0$ and Pauli $\sigma^{1,2,3}$ matrices; while ${(\sigma^{\mu\nu})_a}^b \equiv i (\sigma^\mu_{a\dot{b}} \bar{\sigma}^{\nu\dot{b}b} - \sigma^\nu_{a\dot{b}} \bar{\sigma}^{\mu\dot{b}b})/4$. Finally, $c_{\gamma} \equiv -c_B\cos\theta_W+c_W\sin\theta_W$ is the $\gamma\nu\psi_{3/2}$ coupling.

The contribution of the spin-$\frac32$ singlet fermion to the anomalous magnetic moment is given by \cite{Criado:2021qpd}
\be
    \Delta {a}_\mu^{\psi} = \frac{m_\mu^2 v^2 m_{3/2}^2}{8\pi^2\Lambda^6} \left[ |c_W|^2 f_1(m_{3/2}) + \frac{\textrm{Re} ( c_W^\ast c_{\gamma} )}{\sin(\theta_W)} f_2(m_{3/2}) \right],
    \label{eq:result}
\ee
where the functions $f_1$ and $f_2$ are given by
\be
    f_1=-\frac{13}{27}+\frac{7}{18}\log\left(\frac{\mu^2}{m_{3/2}^2}\right), \qquad f_2=\frac{2}{3}\log\left(\frac{\mu^2}{m_{3/2}^2}\right)\, .
\ee
when $m_{3/2} \gg M_W$, in the $\overline{\rm MS}$-renormalization scheme with a scale $\mu$.
 The contribution is, thus, of order $\Lambda^{-6}$ with the scale $\Lambda$ may be associated with the compositeness scale. 

Eq.~\eqref{eq:result} gives the contribution from $\psi$ to the magnetic moment at a high-energy scale, and its value has to be run down to the scale of the muon mass. Following Ref.~\cite{Aebischer:2021uvt} in which the running and matching from several scales to low energies in the case of the muon dipole moments has been derived, and assuming that $m_{3/2}$ is sufficiently close to the reference value of $250$~{GeV} so that one can fix the renormalization scale $\mu$ to this value, one finds a corrected value given by Eq.~\eqref{eq:result} should be corrected by a factor $0.89$.
%§§§§§§§§§§§§§§§§§§§§§§§§§§§§§§§§§§§§§§§§§§§§§§§§§§§§§§§§§§§§§§§§§§§§§§§§§§§§§§

\section{Numerical results}

\subsection{Spin--1/2 fermions}

\begin{figure}[!h]
\centering
 \hspace{-8mm}
 \includegraphics[width=0.6\linewidth]{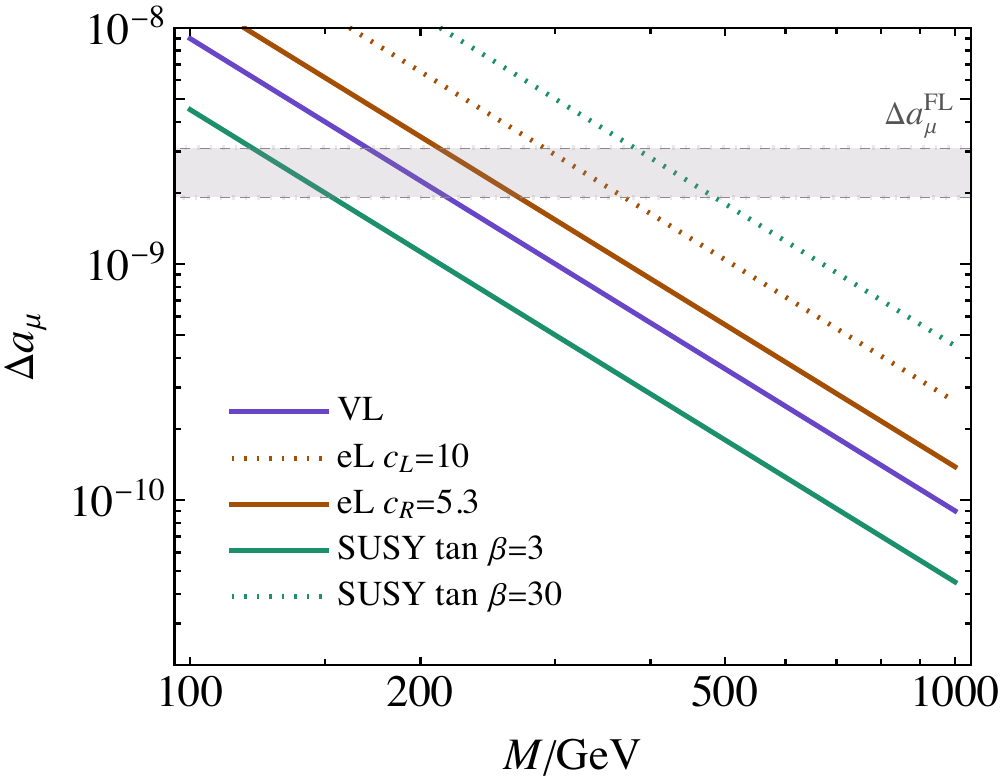}
\caption{
Contributions to the $(g-2)_\mu$ from various spin-$\frac12$ fermions as functions of the corresponding mass scale $M$: vector-like leptons for $\lambda_{LE}\! =\! \overline{\lambda}_{LE}\! =\! 1$ and $M \!=\! \sqrt{M_{L}M_{E}}$ (purple line), excited leptons with $\kappa_{L,R}\!=\! 1$ and $M\!=\!\Lambda$ for the two cases $c_{L} \approx 10$ (dotted red line) and $c_{R} \approx 5.3$ (solid red line) and supersymmetric particles for $M\!=\!\tilde m$ for the two cases of $\tan \beta \!= \!3$ (solid green line) and $\tan \beta \!= \! 30$ (dotted green line). The light grey band shows the $1\sigma$ region of the Fermilab measurement.}
\label{fig3}
\end{figure}

Our numerical results for the three cases of exotic spin-$\frac12$ fermions discussed in the previous subsections are collected in Figure~\ref{fig3} where we present the typical predictions for their contributions to the $(g-2)_\mu$ as a function of their corresponding mass scale. In the case of vector-like fermions, the $M$ scale is defined as $M\! =\! \sqrt{M_{L}M_{E}}$ which, together with the assumption that the Yukawa couplings are simply given by $\lambda_{LE} = \overline{\lambda}_{LE} = 1$, leads to the curve displayed in purple in Fig.~\ref{fig3}. For excited leptons, the scale $M$ is defined in the simplest way as $M\!=\!\Lambda\!= \!m_{\mu^*} \!= \!m_{\nu_\mu^*}$, and we have considered two extreme situations, $\kappa_L=1$ and $\kappa_R=1$, which lead to the coefficient values $c_L=10$ and $c_R=5.3$ respectively. The resulting contributions to $a_\mu$ are presented in the figure in red colors. Finally, in the supersymmetric case, the scale is simply the common mass of the scalar leptons $M= \tilde m$, and we have chosen the values $\tan\beta=3$ and $\tan\beta=30$ to illustrate our results. The resulting typical contributions to $a_\mu$ are shown by the green curves in Fig.~\ref{fig3}. The results of the new Fermilab $(g-2)_\mu$ measurement, including the $\pm 1\sigma$ band, are displayed by the grey band.

A comparison of the predicted results with the new $(g-2)_\mu$ Fermilab measurement indicates that all the considered spin-$\frac12$ new fermions could explain the discrepancy with respect to the SM prediction for new particle masses in the vicinity of a few hundred GeV. In turn, if the latter discrepancy has to be attributed to additional theoretical errors, for instance, the models would be severely constrained by the experiment and, typically, the scale of new physics would be constrained to be above several hundred GeV. 

\subsection{Spin--3/2 leptons}

\begin{figure}[!h]
\centering
\hspace{-8mm}
 \includegraphics[width=0.6\linewidth]{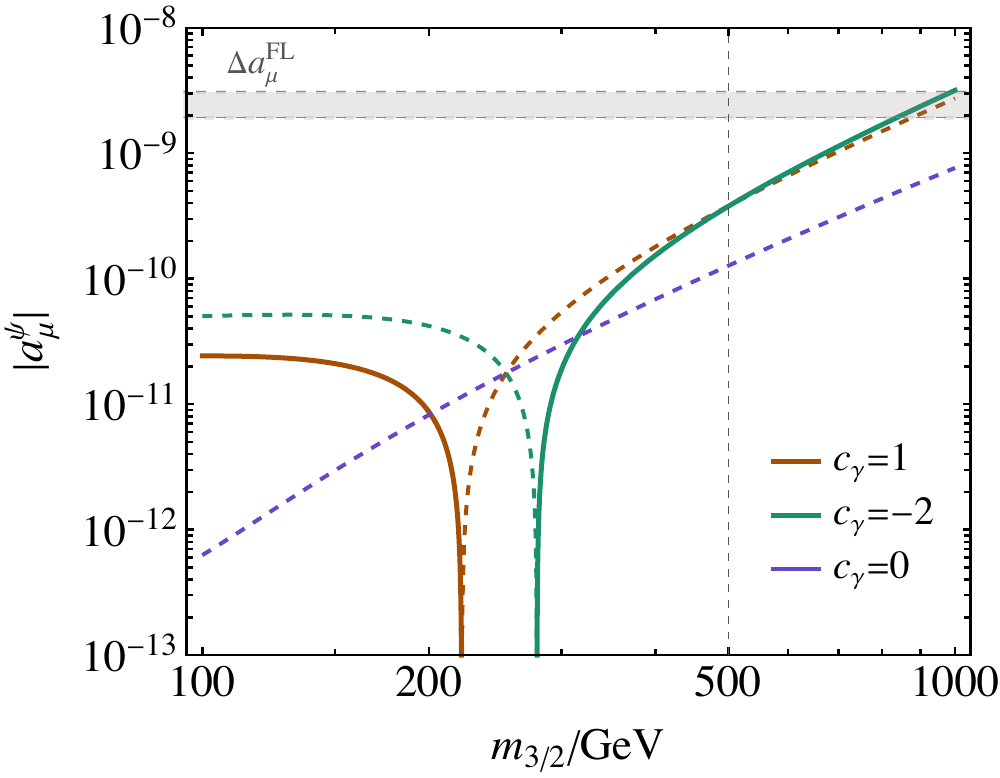} 
\caption{ $a_\mu^\psi$ for different $m_{3/2}$ and $c_{\gamma}$ values while the other parameters are fixed as $\Lambda=500$~GeV, $\mu=250$~GeV and $c_W=1$. The solid and dashed lines correspond to $a_\mu^\psi>0$ and $a_\mu^\psi<0$, respectively. The light grey region represents the $1\sigma$ band of the Fermilab measurement. On the right of the vertical thin dashed line $m_{3/2} > \Lambda$.
}
\label{fig2}
\end{figure}

The contribution to $(g-2)_{\mu}$ from the higher-spin field as a function of its mass $m_{3/2}$ and for different values of the parameter $c_{\gamma}$ is shown in Fig.~\ref{fig2} for a new physics scale $\Lambda=500$~TeV. The results can be roughly summarized in terms of the two mass parameters as
\be\label{a_mu_bound}
    |a_\mu^\psi| \lesssim 2 \times 10^{-11} \left[ \frac{\Lambda}{\TeV} \right]^{-6} \left[ \frac{m_{3/2}}{\TeV} \right]^{2},
\ee
when $c_W, c_\gamma<1$ as expected in the EFT approach. This contribution to $(g-2)_{\mu}$ is consistent with the SM unless the EFT scale is close to the electroweak scale, $\Lambda < 250$ GeV, in which case the validity of the EFT approach starts to be questionable. Also, note that Fig.~\ref{fig2} slightly violates the bound Eq.~\eqref{a_mu_bound} for masses close to the EFT scale. This behaviour is simply an artefact of the large logarithm $\log(m_{3/2}/\mu)$ that is present. In addition, the contribution $a_\mu^\psi$ is negative when $c_\gamma = 0$. A positive $a_\mu^\psi$ value can be obtained for specific values of the model parameters, namely, for sufficiently low $m_{3/2}$ values when $c_\gamma > 0$, or for high enough $m_{3/2}$ values when $c_\gamma < 0$.

As can be seen, for $m_{3/2} \lsim \Lambda=500$ GeV, the spin-3/2 contribution to $(g-2)_{\mu}$ is typically of order  $10^{-10}$--$10^{-11}$, more than an order of magnitude below the experimental sensitivity in the most favourable case. For a particle with such mass and couplings, the production cross section at the LHC in the process $pp \to q\bar q' \to W^*\to \psi_{3/2} \mu +X$, as calculated in Ref.~\cite{Criado:2021itq}, was found to be rather significant, reaching a level of 10~fb at very high masses, a rate that should be sufficient to observe the particle (which could mainly decay into a clear signature consisting, e.g., of a $W$ boson and a muon) at the next LHC runs with expected integrated luminosities of several 100~fb$^{-1}$ to several ab$^{-1}$. 

Nevertheless, one can obtain an anomalous $\psi_{3/2}$ contribution close to the measured $(g-2)_{\mu}$ value if both the effective scale $\Lambda$ and the mass $m_{3/2}$ of the new particle are close to the weak scale, ${O}(300\, {\rm GeV})$. Even for a scale $\Lambda=500$ GeV, the spin-$\frac32$ contribution can still reach the measured value as seen in te figure, if its mass is close to 1 TeV. These values are at the boundary of validity of the EFT. In addition, they should lead to a challenging  $\psi_{3/2}$ production cross section at the LHC

\section{Conclusions}

The new measurement of the anomalous magnetic moment of the muon recently performed at Fermilab has a significant deviation from the prediction in the SM, $4.2\sigma$, which is slightly less than the 5$\sigma$ value traditionally set as the threshold to claim the observation of a new phenomenon. This gives hope that, at last, new physics beyond the SM has been found. This hope is nevertheless tempered by possible additional theoretical uncertainties that have been overlooked and an intense effort would be required in order to settle this crucial issue, hopefully before a new and more precise measurement is released by the experiment. In the meantime, one cannot refrain from interpreting this discrepancy, confront it with various models of new physics beyond the SM and draw the resulting conclusions. 

This is what we have done in this mini-review. We have discussed the contributions of various hypothetical new fermions to the $(g-2)_\mu$ and delineated the scale of their masses and couplings that allows to explain the possible excess compared to the SM expectation. We have considered spin-$\frac12$ new leptons with exotic ${\rm SU(2)_L \times U(1)_Y}$ quantum numbers such as vector-like leptons, excited leptons that are present in composite models and supersymmetric particles, namely the combined contributions of neutralinos/charginos with smuons and their associated sneutrinos. All these scenarios have been widely studied in the past and we simply update the results in the light of the new measurement. However, we have also included a new scenario that has been addressed only recently in Ref.~\cite{Criado:2021qpd}. This is the case of a generic massive isosinglet spin-$\frac32$ fermion in which an EFT approach is used to describe the higher-spin fermion interactions involving only the physical degrees of freedom, thus allowing to calculate in a consistent way the physical observables such as the contributions to the $(g-2)$. 

All these new fermions can give significant contributions to the muon $(g-2)$ which, when confronted with the latest experimental measurement, imply that their masses should be below the TeV scale, if they have to explain the discrepancy from the SM expectation (if this discrepancy with the SM result is indeed real). As shown in the two figures that summarize our results, this implies particles with masses in the few hundred GeV range, which could be observed at the next high-luminosity run of the CERN Large Hadron Collider. If the discrepancy is instead due to additional or overlooked theoretical uncertainties, the new result will impose strong constraints on the masses and couplings of the new spin-$\frac12$ and $\frac32$ particles. 

%%%%%%%%%%%%%%%%%%%%%%%%%%%%%%%%%%%%%%%%%%%%%%%%%%%%%%%%%%%%%%%%%%%%%%

\noindent {\bf Acknowledgement:} This work is supported by the Estonian Research Council grants MOBTTP135, PRG803, MOBTT5, MOBJD323 and MOBTT86, and by the European Union through the European Regional Development Fund CoE program TK133 ``The Dark Side of the Universe." J.C.C. is supported by the STFC under grant ST/P001246/1. A.D. is supported by the Junta de Andalucia through the Talentia Senior program and by the grants A-FQM-211-UGR18, P18-FR-4314 with ERDF.

\bibliographystyle{frontiersinHLTH&FPHY} % for Physics and Mathematics articles

\bibliography{any_spin}

\end{document}